\documentstyle[epsfig,aps,prc]{revtex}

\newcommand{\smfrac}[2]{\mbox{$\frac{#1}{#2}$}}
\tightenlines
\begin{document}

\title{Violation of pseudospin symmetry in nucleon-nucleus scattering:
exact relations
}
\author{H. Leeb and S. Wilmsen}
\address{Institut f\"ur Kernphysik, Technische Universit\"at Wien,
Wiedner Hauptstra\ss e 8-10/142, A-1040 Vienna, Austria} 
\date{\today}
\maketitle

\begin{abstract}
An exact determination of the size of the pseudospin symmetry violating 
part of the nucleon-nucleus scattering amplitude from scattering observables
is presented. The approximation recently used by Ginocchio turns out to
underestimate the violation of pseudospin symmetry.
Nevertheless the conclusion of a modestly broken pseudospin 
symmetry in proton-$^{208}$Pb scattering at $E_L=800$\ MeV remains 
valid.
\end{abstract}
%
%
%
%
%
%
\section{Introduction}
The concept of pseudospin was originally introduced 
\cite{Hecht69,Arima69} to explain the quasidegeneracy of spherical 
shell orbitals with nonrelativistic quantum numbers 
($n_r,\ell ,j=\ell+\frac{1}{2}$) and 
($n_r-1,\ell +2 ,j=\ell +\frac{3}{2}$),
where $n_r$, $\ell $ and $j$ are the single-nucleon radial, orbital
angular momentum and total angular momentum quantum numbers. This 
symmetry holds approximately also for deformed nuclei 
\cite{Ratna73,Draayer84,Zeng91} and even for the case of triaxiality 
\cite{Blokhin97,Beuschel97}. Only recently, Ginocchio and coworkers
\cite{Ginocchio97,Ginocchio98a,Ginocchio98b,Ginocchio99a} pointed out that
the origin of the approximate pseudospin symmetry is an invariance of the
Dirac Hamiltonian with $V_V=-V_S$ under specific $SU(2)$ transformations
\cite{Bell75}. Here, $V_S$ and $V_V$ are the scalar and vector potentials,
respectively. In the non-relativistic limit this leads to a Hamiltonian
which conserves pseudospin, 
\begin{equation}
{\bf \tilde s}=2 \frac{\bf s\cdot p}{p^2}{\bf p} - {\bf s} 
\, ,
\label{eq:stilde}
\end{equation}
where ${\bf s}$ is the spin and ${\bf p}$ is the momentum operator of 
the nucleon. For realistic mean fields there must be at least a weak
pseudospin-symmetry violation because otherwise no bound state exists 
\cite{Ginocchio97}.


Already in 1988 Bowlin et al. \cite{Bowlin88} investigated whether 
the symmetry associated with $V_V=-V_S$ manifests itself also in 
proton-nucleus scattering. They evaluated the analyzing power 
$P(\theta )$ and the spin rotation function $Q(\theta )$ under the 
assumption $V_V=-V_S$, where $\theta $ is the scattering angle. Since 
the experimental data deviate significantly from their prediction they 
concluded that the symmetry is destroyed for low-energy proton scattering; 
only at high energies some remnants might survive. However, recently, 
Ginocchio \cite{Ginocchio99b} revisited this question within the 
scattering formalism in terms of pseudospin. Based on a first-order 
approximation he extracted from experimental proton-$^{208}$Pb scattering 
data at $E_L=800$\ MeV \cite{Fergerson86} the pseudospin-symmetry breaking 
part of the scattering amplitude. Contrary to \cite{Bowlin88} he obtained
at all angles a relatively small pseudospin dependent part of the 
scattering amplitude which confirms the relevance of pseudospin symmetry 
also for proton-nucleus scattering at least at medium energies. 

In the present work we reexamine the question of speudospin symmetry 
in nucleon-nucleus scattering and derive an exact relation for the 
nucleon-nucleus scattering amplitude in terms of scattering observables. 
The exact relationship for the pseudospin symmetry violating part of the 
scattering amplitude differs in an essential way from the first-order 
expression used 
by Ginocchio \cite{Ginocchio99b}. Using the same proton-$^{208}$Pb scattering 
data at $800$\ MeV the exact relationship leads at all measured angles to 
a significantly increased violation of the pseudospin symmetry as compared 
to \cite{Ginocchio99b}. Nevertheless, the size of the violation remains 
within the limits which allow one to consider pseudospin symmetry as a 
relevant symmetry in nucleon-nucleus scattering.

In section II we discuss briefly the basic relations between the 
scattering observables and the scattering amplitude of nucleon-nucleus 
scattering within the standard formalism. Introducing the proper 
transformation to a pseudospin representation we present in section III 
exact relations for the pseudospin-symmetry violating part of the 
scattering amplitude. A first application of the exact
relations is given in section IV where we consider the example of 
proton-$^{208}$Pb scattering data at $E_L=800$\ MeV. Concluding 
remarks are given in section V.

\section{Spin dependent scattering amplitude}

The scattering amplitude, $f$, for the elastic scattering of a nucleon
with momentum $k$ on a spin zero target is given by \cite{Feshbach92},
\begin{equation}
f = A(k,\theta ) + 
B(k,\theta ) \mbox{\boldmath $\sigma $}{\cdot {\bf \hat n}}
\, .
\label{f}
\end{equation}
Here \mbox{\boldmath $\sigma $} is the vector formed by the Pauli matrices,
${\bf \hat n}$ is the unit vector perpendicular to the scattering
plane and $\theta $ is the scattering angle. The complex-valued functions 
$A(k,\theta )$ and $B(k,\theta )$ are the spin-independent and the 
spin-dependent parts of the scattering amplitude. Both are not fully 
accessible to measurement. 

The observables in nucleon-nucleus scattering are related to intensities
and can be described in terms of the amplitudes $A(k,\theta )$ and 
$B(k,\theta )$.
In the case of the scattering of a spin-\smfrac{1}{2} particle by a 
spinless target the observables are the differential cross section 
\begin{equation}
\frac{d\sigma }{d\Omega }(k,\theta ) = 
|A(k,\theta )|^2 + |B(k,\theta )|^2
\, ,
\label{ds}
\end{equation}
the polarization
\begin{equation}
P(k,\theta ) = 
\frac{B(k,\theta )A^*(k,\theta )+B^*(k,\theta )A(k,\theta )}
{|A(k,\theta )|^2 + |B(k,\theta )|^2} \, ,
\label{P}
\end{equation}
and the spin rotation function
\begin{equation}
Q(k,\theta ) = \mbox{i} 
\frac{B(k,\theta )A^*(k,\theta )-B^*(k,\theta )A(k,\theta )}
{|A(k,\theta )|^2 + |B(k,\theta )|^2} \, .
\label{Q}
\end{equation}
From Eqs. (\ref{P}) and (\ref{Q}) follows $P^2+Q^2 \leq 1$.

The extraction of the full scattering amplitude (moduli and phases of 
$A(k,\theta )$ and $B(k,\theta )$) from measurements is a very challenging
question in quantum mechanics and intimately related to the longstanding 
{\em phase problem} in structure physics. Here, only a partial solution 
of this problem is required because our interest is limited to the ratio 
$|R_s(k,\theta )|=|B(k,\theta )|/|A(k,\theta )|$ which is a measure of the 
strength of the spin-dependent interaction. Combining 
Eqs. (\ref{P}) and (\ref{Q}) provides a phase relation between the 
amplitudes,
\begin{equation}
\sqrt{\frac{P - \mbox{i} Q}{P + \mbox{i} Q}} = \exp \mbox{i}(\phi_B - \phi_A)
\, ,
\label{phi}
\end{equation}
where $\phi_A$ and $\phi_B$ are the phases of the amplitudes $A$
and $B$, respectively. Using this result either in Eq. (\ref{P}) or 
(\ref{Q}) leads to a quadratic equation for the ratio of the moduli,
\begin{equation}
1 - \frac{2}{\sqrt{P^2+Q^2}} |R_s| + |R_s|^2 = 0
\, ,
\label{quad}
\end{equation}
which has two solutions. Using the condition that $P=Q=0$ implies $|B|=0$ 
one can immediately select the physical solution ($|A|\neq 0$),
\begin{equation}
|R_s| = \frac{|B|}{|A|}=\frac{1-\sqrt{1-P^2-Q^2}}{\sqrt{P^2+Q^2}}
=\frac{\sqrt{P^2+Q^2}}{1+\sqrt{1-P^2-Q^2}}
\, .
\label{absRs}
\end{equation}
This result together with Eq. (\ref{phi}) yields also the ratio of the
amplitudes,
\begin{equation}
R_s = \frac{B}{A} = \frac{P - \mbox{i} Q}{1+\sqrt{1-P^2-Q^2}}
\, .
\label{Rs}
\end{equation}
These relations are exact at all scattering angles and energies.

For comparison with the recent work of Ginocchio \cite{Ginocchio99b} we
evaluate $|R_s|^2$ from Eq. (\ref{Rs}),
\begin{equation}
|R_s|^2 = C_s \left[ \left( \frac{P}{2} \right)^2 + 
\left( \frac{Q}{2} \right)^2 \right] 
\label{absRs2}
\end{equation}
with
\begin{equation}
C_s =\frac{4}{2+2\sqrt{1-P^2-Q^2}-P^2-Q^2} \geq 1
\, .
\label{corrs}
\end{equation}
The second factor (square brackets) on the right-hand side of
Eq. (\ref{absRs2}) is the first-order expression used in \cite{Ginocchio99b}. 
It is obvious from the correction factor $C_s$ that the first-order 
approximation systematically underestimates the quantity $|R_s|^2$.

Eqs. (\ref{phi}) and (\ref{Rs}) represent important pieces for the 
proper determination of the scattering amplitude from nucleon-nucleus 
scattering observables. In addition one can derive from Eq. (\ref{ds})
with the use of Eq. (\ref{Rs}) the relationship,
\begin{equation}
|A|^2 = \frac{d\sigma }{d\Omega } 
\left( 1 - \frac{P^2+Q^2}{2+2\sqrt{1-P^2-Q^2}}\right)
\, .  
\label{A}
\end{equation}
Combining Eqs. (\ref{phi}), (\ref{Rs}) and (\ref{A}) one can determine
the full scattering amplitude up to an overall phase. Thus we have 
reduced the determination of the scattering amplitude in a system with
spin $\frac{1}{2}$ to a problem similar to that with spin zero. 

\section{Violation of pseudospin symmetry}

The closed-form expression (\ref{Rs}) for the spin-dependent term 
indicates the possibility of deriving an exact expression for 
the violation of the pseudospin symmetry. For this purpose the concept
of pseudospin must be introduced in the formalism of nucleon-nucleus
scattering. Specifically, the partial-wave expansion of the scattering
amplitude must be performed in terms of the pseudo-orbital angular 
momentum,
\begin{eqnarray}
\tilde \ell & = & \ell + 1\, , \mbox{ for } j = \ell + 1/2\, ,
\nonumber \\
\tilde \ell & = & \ell - 1\, , \mbox{ for } j = \ell - 1/2\, .
\end{eqnarray}
Accordingly, the partial-wave S-matrix elements $S_{\ell ,j}$ must be 
defined for pseudo-orbital angular momentum ($\tilde S_{\ell ,j}$) as
\begin{equation}
\tilde S_{\tilde \ell,j=\tilde \ell -1/2} = 
S_{\tilde \ell -1,j=\tilde \ell -1/2}\, , \quad 
\tilde S_{\tilde \ell,j=\tilde \ell +1/2} = 
S_{\tilde \ell +1,j=\tilde \ell +1/2}\, .
\end{equation}
As shown by Ginocchio \cite{Ginocchio99b} the scattering amplitudes 
$\tilde A$ and $\tilde B$ are related to $A$ and $B$ by a unitary 
transformation
\begin{equation}
\left( \begin{array}{c}
\tilde A \\ \tilde B \end{array} \right) =
\left( \begin{array}{cc} 
\cos (\theta ) & \mbox{i}\sin (\theta ) \\
\mbox{i}\sin (\theta ) & \cos (\theta ) 
\end{array} \right) \left( \begin{array}{c}
A \\ B \end{array} \right) \, .
\label{trafo}
\end{equation}

Using the transformed amplitudes the scattering observables are then
given by
\begin{equation}
\frac{d\sigma }{d\Omega } = |\tilde A|^2 + |\tilde B|^2
\, ,
\label{sigmas}
\end{equation}
\begin{equation}
P = \frac{\tilde B \tilde A^* + \tilde B^* \tilde A}
{|\tilde A|^2 + |\tilde B|^2}
\, ,
\label{Ps}
\end{equation}
\begin{equation}
Q = \frac{\sin (2\theta ) \left[ |\tilde A|^2 - |\tilde B|^2 \right]
+\mbox{i} \cos (2\theta ) 
\left[ \tilde B \tilde A^* - \tilde B^* \tilde A\right]}
{|\tilde A|^2 + |\tilde B|^2}
\, .
\label{Qs}
\end{equation}
In the limit of pseudospin symmetry the amplitude $\tilde B$ vanishes
and consequently $P=0$ and $Q=\sin (2\theta )$ \cite{Bowlin88}. Therefore
the ratio $R_{ps}=\tilde B/\tilde A$ is a measure of the violation of
pseudospin symmetry in nucleon-nucleus scattering.

With the transformation (\ref{trafo}) it is straightforward to 
express $R_{ps}$ in terms of $R_s$,
\begin{equation}
R_{ps} = \frac{\tilde B}{\tilde A} =
\frac{\mbox{i} \tan (\theta ) +  R_s}
{1 + \mbox{i} \tan (\theta ) R_s}
\label{Rps}
\end{equation}
and
\begin{equation}
|R_{ps}|^2 = \frac{\tan^2(\theta ) - Q\tan (\theta ) 
+ |R_s|^2(1-Q\tan (\theta ))}
{1+Q\tan (\theta ) + |R_s|^2 (\tan^2 (\theta ) + Q\tan (\theta ))}
\, .
\label{absRps2}
\end{equation}
Substitution of Eq. (\ref{Rs}) in Eq. (\ref{Rps}) yields an exact 
relation for the violation of pseudospin symmetry in terms of scattering
observables,
\begin{equation}
R_{ps} = \frac{P+\mbox{i}\left[\left( 1+\sqrt{1-P^2-Q^2}\right) 
\tan (\theta ) - Q\right]}{\left[
\left(1+\sqrt{1-P^2-Q^2}\right) + Q\tan (\theta )\right]
+\mbox{i}P\tan (\theta )}
\, .
\end{equation}
For comparison with the recent work of Ginocchio \cite{Ginocchio99b} we
evaluate $|R_{ps}|^2$ from Eq. (\ref{Qs}). By straightforward algebraic
manipulations one obtains
\begin{equation}
|R_{ps}|^2 = C_{ps} \left[ \left(\frac{P}{2}\right)^2 
+\left(\frac{Q-\sin (2\theta )}{2\cos (2\theta )}\right)^2\right] 
\label{Rps2}
\end{equation}
with
\begin{equation}
C_{ps} = \frac{\left(1+|R_{ps}|^2\right)^2}{1 + |R_{ps}|^2\tan^2 (2\theta )
+\mbox{i} \left(R_{ps}^* - R_{ps} \right) \tan (2\theta )}
\, .
\label{corrps}
\end{equation}
The factor in square brackets on the right-hand side of Eq. (\ref{Rps2}) 
corresponds to the first-order expression used in \cite{Ginocchio99b}
while $C_{ps}$ is a correction factor. Eqs. (\ref{Rps2}) and
(\ref{corrps}) are not the best suited for the evaluation of 
$|R_{ps}|^2$, however, they demonstrate clearly that the actual value of
the pseudospin symmetry breaking amplitude may deviate significantly
from the first-order approximation.  

\section{Example of proton-nucleus scattering}

The relations derived in section II and III can be directly used for 
an analysis of proton-nucleus scattering data, where accurate measurements
of the analyzing power $P(\theta )$ and the spin rotation function 
$Q(\theta )$ are available. Specifically, we 
consider analyzing power \cite{Hoffmann78} and spin rotation function 
measurements \cite{Fergerson86} for proton-$^{208}$Pb scattering at 
$E_L=800$\ MeV and evaluate the ratios $|R_s|^2$ and $|R_{ps}|^2$.

%
The violation of pseudospin symmetry can be read off from Fig. 1, where
the ratio $|R_{ps}|^2$ is displayed as a function of the scattering angle.
In Fig. 1 the ratios are shown only at those scattering angles where 
measured values of the spin rotation function \cite{Fergerson86} are 
available. The corresponding values of the analyzing power have been
obtained by linear interpolation of the more complete data set given by 
\cite{Hoffmann78}. The shown uncertainties result from a linear error
propagation of the given experimental uncertainties in $P(\theta )$ 
and $Q(\theta )$. For comparison the results of \cite{Ginocchio99b}
are also displayed. It is obvious from Fig. 1 that the first-order 
approximation underestimates the pseudospin symmetry breaking 
part of the scattering amplitude at all scattering angles, particularly 
at the highest available ones. Furthermore it is interesting to note 
that the uncertainties obtained with the use of the exact relations are 
consistently larger than those obtained in first-order approximation. 

In Fig. 2 the corresponding results are shown for the 
ratio $|R_s|^2$ characterizing the spin dependent part of the
scattering amplitude. As already seen from the exact formulation
(\ref{absRs2}) the first-order approximation underestimates 
systematically the spin dependent part. The difference depends only 
on the absolute size of $P^2+Q^2$ and amounts to a factor $2$ at
$\theta =15$ degrees.
%

The application of Eqs. (\ref{phi}), (\ref{Rs}) and (\ref{A}) to 
scattering data is straightforward and yields $|A(k,\theta )|$, 
$|B(k,\theta )|$ and $\phi_B(k,\theta )-\phi_A(k,\theta )$. Up to a 
common phase, therefore, the scattering amplitude can be extracted from
the scattering observables.

\section{Conclusions}

We have derived closed-form expressions for the relationship between
scattering amplitude and observables in nucleon-nucleus scattering.
The scattering amplitudes can be determined, up to a common phase
factor, from measured differential cross section, analyzing power and 
spin rotation data. Hence the problem of extracting the scattering 
amplitudes from nucleon-nucleus scattering data becomes mathematically
similar to that for spinless particles.

Extending these relations we derived a closed-form expression for the 
ratio of the pseudospin dependent and independent amplitudes. Thus 
we obtain an exact measure of the violation of pseudospin symmetry.
Comparison with the first-order expression of \cite{Ginocchio99b}
exhibits significant deviations. Using experimental proton-$^{208}$Pb 
data at $E_L=800$\ MeV of \cite{Fergerson86,Hoffmann78} we find an 
increased violation of pseudospin symmetry in the whole available 
angular range as compared to the results of \cite{Ginocchio99b}. The
increment is up to $40$\% so that $|R_{ps}|^2$ reaches values of about
$0.15$ at $\theta = 15$\ degrees. Nevertheless it remains within the
limits which allow one to consider the pseudospin symmetry to be a relevant 
concept in nucleon-nucleus scattering at intermediate energies. 
The ratio of of pseudospin to spin dependence is actually smaller than
estimated in \cite{Ginocchio99b}, e.g. at $\theta = 15$ degrees 
$|R_{ps}|^2/|R_s|^2 \approx 1/2.4$ instead of approximately $1/2$ in
\cite{Ginocchio99b}. This difference is a direct consequence of the
remarkable spin-dependence of the scattering amplitude which limits the
reliability of the estimation of $|R_s|^2$ within perturbation theory.
At lower energies increased violation of pseudospin symmetry may occur
\cite{Bowlin88}. Whether this is true has to be proved by studying the 
energy dependence of $|R_{ps}|^2$ for several nuclei.

\acknowledgements
The authors thank Prof. Dr. G.W. Hoffmann and Prof. Dr. L. Ray for making 
available the experimental data in tabular form and Prof. Dr. R. Lipperheide 
for fruitful discussions and a careful reading of the manuscript.

%
%
%
%
%
%
\vfill
\newpage

\section*{Figure Captions}

{\bf Figure 1}\\
The angular dependence of the ratio 
$|R_{ps}|^2$
evaluated
from proton-$^{208}$Pb scattering data at $E_L=800$\  MeV.
The solid error bars are obtained by the exact formulation 
(\ref{absRps2}), the dashed error bars are the results of the 
first-order approximation. The lines connecting the datapoints are 
drawn as a guidance for the eye.

\vspace{0.5cm}
{\bf Figure 2}\\
The angular dependence of the ratio 
$|R_{s}|^2$
evaluated
from proton-$^{208}$Pb scattering data at $E_L=800$\  MeV. 
The solid error bars are obtained by the exact 
formulation (\ref{absRps2}), the dashed error bars are the results 
of the first-order approximation.
The lines connecting the datapoints are drawn as a guidance for the 
eye.



\newpage

\begin{center}

\begin{figure}[htb]
\centerline{\epsfig{file=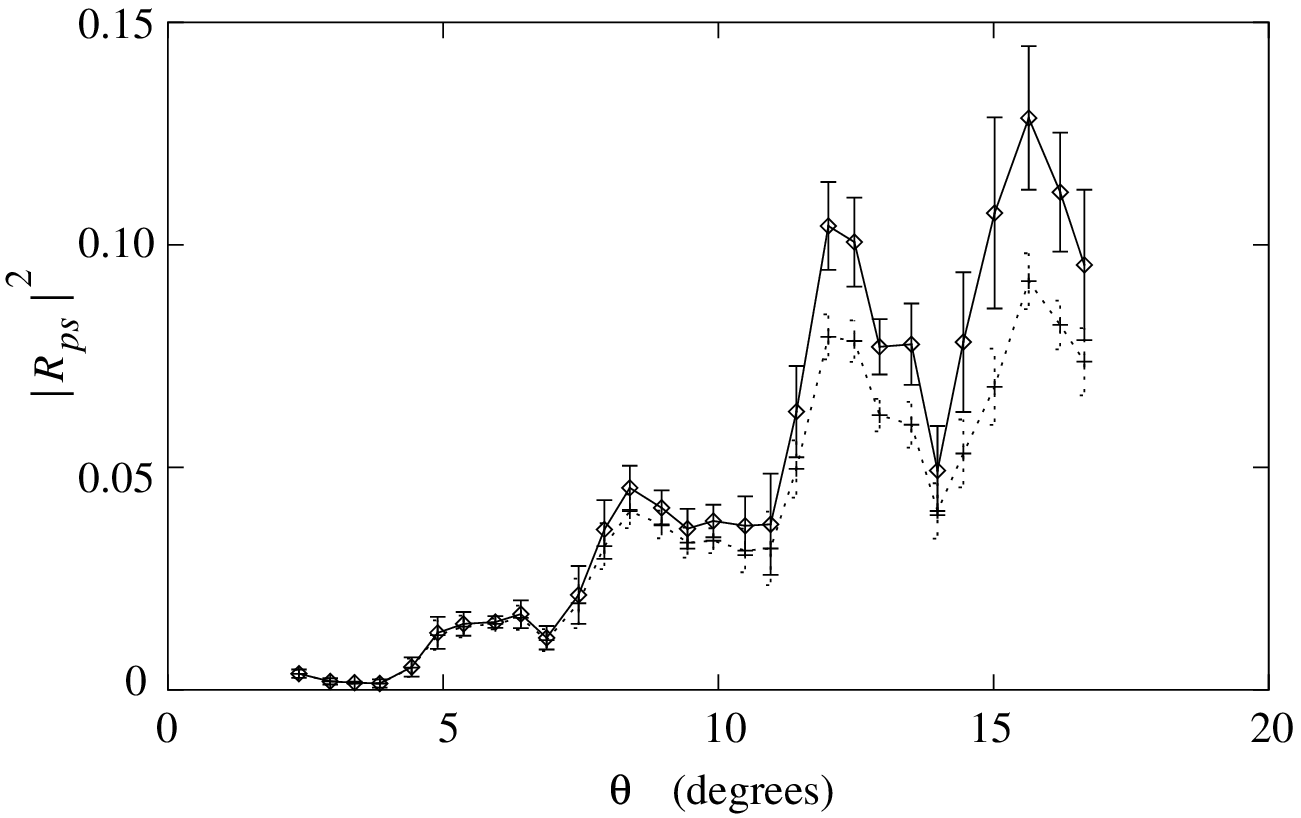,width=16.0cm}}
\end{figure}

\vspace{3cm}

{\large {\bf Figure 1}}

\newpage

\begin{figure}[htb]
\centerline{\epsfig{file=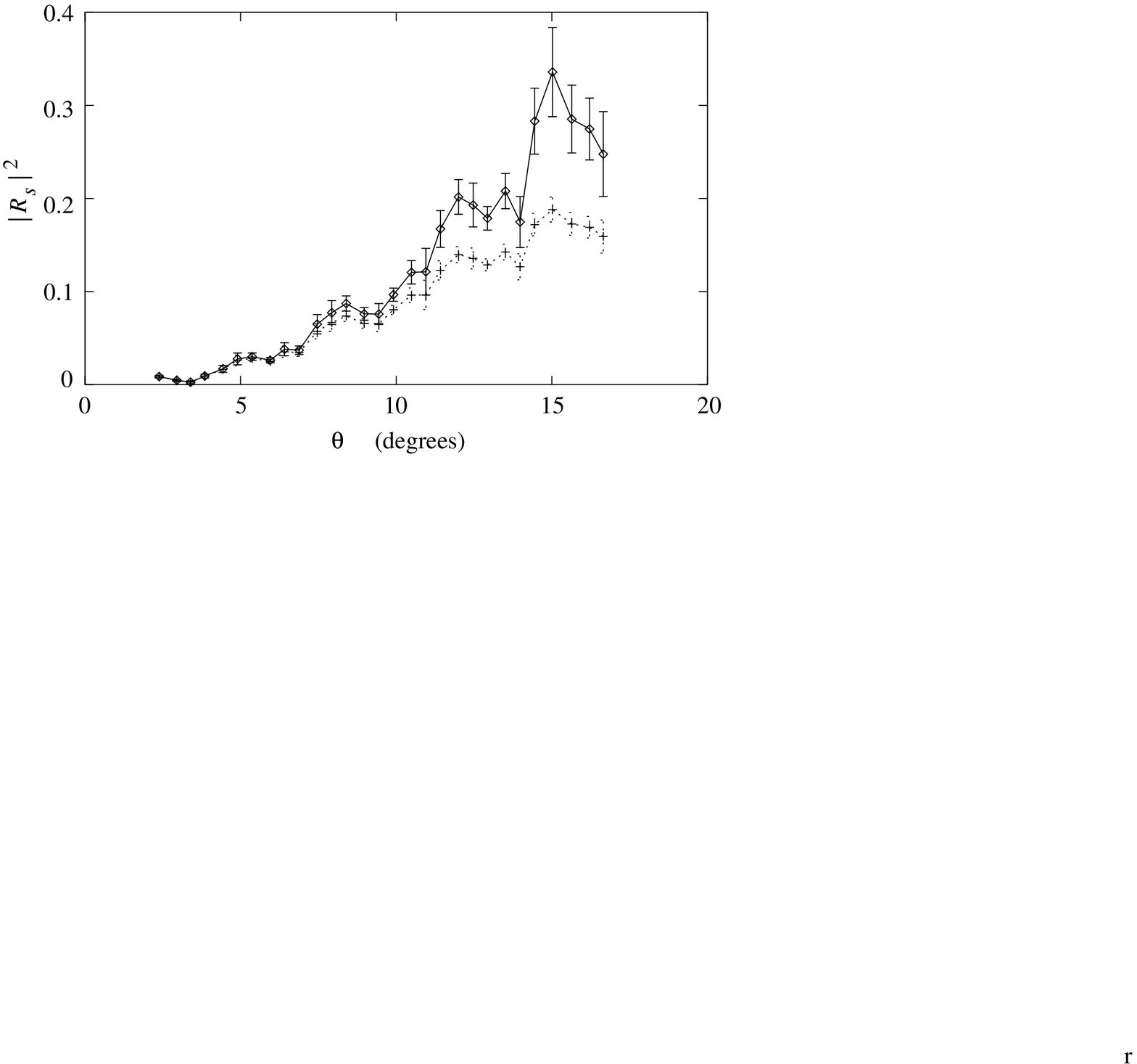,width=16.0cm}}
\end{figure}

\vspace{3cm}

{\large {\bf Figure 2}} 
\end{center}

%

\end{document}